\documentclass[11pt,a4paper]{article}
\usepackage{bm,jcappub,Macro,tabularx}
\usepackage{url}
\usepackage{hyperref}
\usepackage{url}
\usepackage{epstopdf}
\usepackage{enumitem}
\usepackage{colortbl}
\usepackage{color}

\def\xt{({\bf x},\tau)}
\def\xti{({\bf x},\tau_i)}
\def\ti{\tau_i}
\def\cG{{\cal G}}

\def\la{\langle}
\def\ra{\rangle}

\def\be{\begin{equation}}
\def\ee{\end{equation}}
\def\ben{\begin{eqnarray}}
\def\een{\end{eqnarray}}
\def\bes{\begin{subequations}}
\def\ees{\end{subequations}}
\def\nn{\nonumber}

\def\bk{{\bf k}}

\def\bk{{\bf k}}

\def\bx{{\bf x}}
\def\2p{{(2\pi)^2}}

\def\be{\begin{equation}}
\def\ee{\end{equation}}
\def\beq{\begin{equation}}
\def\eeq{\end{equation}}
\def\ben{\begin{eqnarray}}
\def\een{\end{eqnarray}}

\def\nn{{\nonumber}}

\newcommand{\beqa}{\begin{eqnarray}}
\newcommand{\eeqa}{\end{eqnarray}}

\def\ikap0{{\cal J}_{\theta_0}(r)}

\def\one1{\langle \kappa_{(i)}\kappa_{(j)} \rangle}
\def\one{{[\bar \xi^{(ij)}]}}

\def\ba{\begin{eqnarray}}
\def\ea{\end{eqnarray}}

\def\bk{{\bf k}}
\def\bq{{\bf q}}

\title{Symmetries, Invariants and Generating Functions: Higher-order Statistics of Biased Tracers}
\author{Dipak Munshi}
\affiliation{Astronomy Centre, School of Mathematical and Physical Sciences,\\ University of Sussex, Brighton BN1 9QH, U.K.}
\emailAdd{D.Munshi@sussex.ac.uk}
\abstract{Gravitationally collapsed objects are known to be biased tracers of an underlying density contrast.
Using symmetry arguments, generalised biasing schemes have recently been developed to
relate the halo density contrast $\delta_h$ with the underlying density contrast $\delta$, divergence of velocity $\theta$ 
and their higher-order derivatives. This is done by constructing invariants such as $s, t, \psi,\eta$.
We show how the generating function formalism in Eulerian standard perturbation theory (SPT) can be used to show that many of the
additional terms based on extended Galilean and Lifshitz symmetry actually do {\em not} 
make any contribution to the higher-order statistics of biased tracers.
Other terms can also be drastically simplified allowing us to write the vertices associated with
$\delta_h$ in terms of the vertices of $\delta$ and $\theta$, the higher-order derivatives and the bias coefficients.
We also compute the cumulant correlators (CCs) for two different tracer populations. 
These perturbative results are valid for {\em tree-level} contributions but at an {\em arbitrary order}. 
We also take into account the stochastic nature bias in our analysis.
Extending previous results of a local polynomial model of bias, we express the one-point cumulants ${\cal S}_N$ and their two-point counterparts,
the CCs i.e. ${\cal C}_{pq}$, of biased tracers in terms of that of their underlying density contrast counterparts.
As a by-product of our calculation we
also discuss the results using approximations based on Lagrangian perturbation theory (LPT).}
\keywords{Cosmology, Large-Scale Structure, Perturbation Theory} 
\begin{document}  
\maketitle
\bigskip
\bigskip
{\it ``Why do you want to know his bias? Form your own bias!''  R.P.Feynman}
%
\section{Introduction}
\label{sec:intro}

Recently completed cosmic microwave background (CMB) surveys e.g. the 
Planck surveyor\footnote{Planck: \href{http://www.cosmos.esa.int/web/planck/}
{\tt  http://www.cosmos.esa.int/web/planck/}}, have provided us with a robust cosmological framework 
that will allow us to investigate the physics beyond the {\em Standard Model} of cosmology.
Next generation surveys are mapping the entire CMB sky with higher resolution and accuracy
e.g. ACT\footnote{ACT: \href{http://www.physics.princeton.edu/act/}{\tt http://www.physics.princeton.edu/act/}} and
SPT\footnote{SPT: \href{http://pole.uchicago.edu/}{\tt http://pole.uchicago.edu}} to answer many of the
questions relevant to structure formation in the low redshift Universe.
In addition, the ongoing and future large scale surveys will map the sky with ever increasing precision
(BOSS\footnote{Baryon Oscillator Spectroscopic Survey: \href{http://www.sdss3.org/surveys/boss.php}{\tt http://www.sdss3.org/surveys/boss.php}}
\citep{EW},
WiggleZ\footnote{WiggleZ Survey : \href{http://wigglez.swin.edu.au/}{\tt http://wigglez.swin.edu.au/}}
\citep{DJA},
DES\footnote{Dark Energy Survey: \href{http://www.darkenergysurvey.org/}{\tt http://www.darkenergysurvey.org/}}
\citep{DES},
EUCLID\footnote{EUCLID: \href{http://www.euclid-ec.org/}{\tt http://www.euclid-ec.org/}}
\citep{LAA}). These surveys will provide a glimpse of physics beyond the standard model.
On one hand, they will check any departure from general relativity (GR) on cosmological scales,
on the other, they will also provide an estimate of the sum of the neutrino mass \citep{review2}. 
The galaxies, however, are known to be biased tracers of underlying dark matter distribution \citep{Desjacques}.
To achieve the full potential of the future surveys, parametrization and
understanding of bias is of utmost importance.

Initial models of bias were linear and local relations
between the tracer's density contrast $\delta_h$ and the underlying matter density contrast $\delta$:  $\delta_h=b_1\delta$.
However, early simulations e.g. \citep{CenOstriker} pointed to a more complex nature of bias that can
be nonlinear and non-local. It was also realized that bias can also be stochastic. 
On the theoretical side, both perturbative and non-perturbative models started to emerge.
The non-perturbative theories based on peak approach were developed in \citep{BBKS};
these theories can mimic many of the properties of galaxy bias successfully. 
Halo based approaches were developed which remain an important tool for making
analytical predictions \citep{CooraySheth}.

Many recent approaches have also seen development of formal perturbative
approaches that extend and put known results on a solid foundation \citep{Matsubara,Carlson,Vlah}.   
The effective field theory based approaches were also developed \citep{Effective_bias,Angulo}.
Indeed, the polynomial model for bias developed in \citep{FryGaztanga} was valid for large smoothing scales 
and lacks the non-local terms that are generated due to gravitational clustering.
Using just second-order standard perturbation theory (SPT) it has been shown that gravitational
evolution is responsible for generating tidal interactions which are non-local in the density field \citep{Fry,Goroff,Bouchet}.
Similar non-local contribution is also expected in the clustering of halos and was studied in detail in several
publications \citep{Catelan1,Catelan2}. In \citep{Roy} a more generic scheme for bias was developed based on
symmetries inherent in the dynamical equations.  
In the nonlinear regime gravity-induced bias has been studied in the context of the
hierarchical ansatz (HA) \citep{bias_letter, BernardeauSchaeffer}.
It has recently been pointed out that non-local bias can mimic scale-dependent 
suppression of growth of perturbation in cosmologies with massive neutrinos  
\citep{Saito,Audren,Gong,Adreu,Beutler,Marlina}.
The non-local bias is relatively small compared to the linear bias
but future large-scale surveys will be sensitive to them.
Large-scale numerical simulations have been employed to investigate their
effect on the clustering of halos \citep{sim1,sim2,sim3, LoVerde}. It was shown that
scale-dependent bias in the power spectrum of dark matter halo is
degenerate with the signatures left by the various non-local bias terms.
The degeneracies present in the characterization of the bias can be broken by
using higher-order statistics of biased tracers. Going beyond the usual power spectrum analysis, 
in this paper we compute the higher-order statistics for a generic biasing scheme
which can be nonlinear, non-local and stochastic. The additional
terms in the perturbative description of bias stem from symmetry considerations.
The inherent symmetries in the dynamical
equations predict invariant quantities \citep{ChecnScocSheth,Baldauf,Kehagias} in the perturbative
expansion of bias. 

A generating function based approach in the perturbative regime was introduced in \citep{Ber92}.
It provides a powerful framework to analyse the higher-order statistics of cosmological fields.
We use this formalism along with the
functional relationship dictated by a biasing scheme of $\delta_h$ with $\delta$ and $\theta$ to compute the
tree-level vertices of $\delta_h$ as a function of those of $\delta$ and $\theta$.
These derivations are valid in the perturbative regime.
We derive the formal relations in the presence of nonlinear, non-local and higher-derivative terms.
We also include a stochastic noise that originates from our lack of 
knowledge of the fundamental physics related to the galaxy formation process.
We use these expressions to decide which terms in these expressions
do not contribute at any order. The results from the generating function formalism 
are next used to express the cumulants and cumulant-correlators of the biased tracers.
Our aim is to  extend the results presented in \citep{Munshi_IBIT} to more general biasing schemes. 

This paper is organised as follows. In 
\textsection{\ref{sec:gen}} we review the generating function approach.
In \textsection{\ref{sec:non}} we consider a family of generalized bias models and use the generating function to analyse them.
\textsection{\ref{sec:bias}} is devoted to discussion of a non-local bias.
\textsection{\ref{sec:cumu}} is devoted to the discussion of cumulants and cumulant correlators (CCs).
Finally, our conclusions are presented in \textsection{\ref{sec:conclu}}. Some of the details of our derivations
are relegated to the two appendices.
  
\section{Generating Functions}
\label{sec:gen}
%
The generating function formalism is often used to compute the cumulants and the CCs of cosmological fields.
Our aim here is to provide a very brief review of the generating function formalism developed in \citep{Ber92} 
(also see \citep{MSS}) to 
construct the CCs of the biased tracers (halos or peaks) and express them in terms of the statistics
of underlying mass distribution. The results are relevant for the perturbative regime.
The $n$-th order of perturbative expansion of an arbitrary field $F$ defined as
$F^{(n)}$ with respect to $\delta$ is defined as follows:
\ben
\label{eq:gen}
\la {F}^{(n)}\ra_c = 
{{\int \la {\rm F}^{(n)}({\bf x}, a)\delta^{(1)}({\bf x}_1,a)\cdots \delta^{(1)}({\bf x}_n,a)\ra_c\; d^3{\bf x}\,d^3{\bf x}_1 \cdots d^3{\bf x}_n}
\over 
({\int \la\delta^{(1)}({\bf x},a)\delta^{(1)}({\bf x^{\prime}},a) \ra 
d^3{\bf x} d^3{\bf x}^{\prime}})^n}.
\een
Here, $\delta^{(1)}({\bf x},a)$ is the
linear approximation for $\delta({\bf x},a)$ at a comoving position ${\bf x}$ and $a(t)$ is the scale factor of the Universe. Only
connected diagrams are taken into
account, which explains the subscript $c$. Throughout, we will assume that the
initial density contrast $\delta$ is Gaussian, though it is possible to incorporate non-Gaussian initial condition.
The generating function ${\cal G}_{F}(\tau_s)$ for the vertices for any random field $F({\bf x},a)$ is given by:
\ben
{\cal G}_{F}(\tau_s) = \sum^{\infty}_{n=1} {\la {F}^{(n)} \ra_c \over n!}\tau_s^n.
\een

For two arbitrary fields $A(\bx,a)$ and $B(\bx,a)$, we have the following properties 
for the generating functions \citep{Ber92}:
\bes
\ben
\label{eq:gen_1}
&& {\cal G}_{A+B}(\tau_s) = {\cal G}_A(\tau_s)+{\cal G}_B(\tau_s); \quad\\
&& {\cal G}_{AB}(\tau_s) = {\cal G}_A(\tau_s){\cal G}_B(\tau_s); \quad\\
\label{eq:gen_2}
&&{\cal G}_{\nabla_i A \; \nabla_i B}(\tau_s) = 0; \quad\\
\label{eq:gen_3}
&& {\cal G}_{\nabla_i\nabla_j A \;\nabla_j\nabla_i B}(\tau_s) = {1 \over 3}{\cal G}_{\nabla^2 A\nabla^2 B}(\tau_s).
\label{eq:gen_n}
\een
\ees
We will denote the generating function of the density contrast $\delta$ by  ${\cal G}_{\delta}(\tau_s) = \sum^{\infty}_{n=1}\, {\nu_n/n!}\, \tau^n_s$
where $\nu_n \equiv \la \delta^{(n)} \ra_c$. We will also need the divergence of velocity $\theta$ (to be defined later),
for which, the generating function will be denoted as ${\cal G}_\theta(\tau_s)\equiv  \sum^{\infty}_{n=1}\, {\mu_n/n!}\, \tau^n_s$ 
with $\mu_n \equiv \la \theta^{(n)} \ra_c$.

The generating function formalism was  originally introduced in \citep{BaS89} and later
exploited in many publications including in \citep{Ber92,Ber94} to compute the lower order cumulants and cumulant correlators \citep{francis}
by linking the generating function of tree-level amplitudes directly with the dynamical
equations of a self-gravitating collisionless system \citep{review}. 
The resulting expressions will be applied to understand halo clustering in \textsection{\ref{sec:non}}.
Notice the bias of overdense regions has been studied using
the generating function formalism in  \citep{francis}.
%

\section{Bias and Biased Tracers}
\label{sec:non}
%
The idea of nonlocal bias has been investigated in great detail in the past by many author. Starting from ref.\citep{Roy} the idea was developed 
further in \citep{ChecnScocSheth,Baldauf}. More recently the results were extended to third order in perturbation theory in ref.\citep{halo}.
The idea behind these studies is to probe the statistics of proto-halos which preserve their identity and
their number density is conserved.  It is assumed that, though their shape and topology may change, the center of mass
of these proto-halos follow a well defined trajectory and their statistics can studied using perturbative techniques. 
We will consider the matter dominated case, ${\bf x}$ is the spatial comoving coordinate and $\tau$
the conformal time $\tau = \int {dt/a(t)}$. The associated Hubble parameter is ${\cal H} = {d\ln a(t)/d\tau}$.
We will also define the divergence of velocity as $\theta = \partial_i{v^i}$ where $v^i = dx^i/d\tau$.
The Euler, continuity and Poisson equations describe the gravitational clustering of a collisionless system in
the hydrodynamic limit: 
\bes
\ben
&& {\partial \delta \over \partial \tau} + \nabla_i[(1+\delta) v^i]=0; \\
&& {\partial v^i \over \partial \tau} + {\cal H} v^i + v^j\nabla_j v^i = -\nabla^i \phi;\\
&& \nabla^2\phi = {3 \over 2}\Omega_{\rm M}{\cal H}^2 \delta.
\een
\ees
In terms of generating functions for $\delta$ and $\theta$ denoted as 
${\cal G}_\delta = \sum ({\nu_n/n!})\tau_s^n$ and ${\cal G}_\theta = \sum ({\mu_n/n!})\tau_s^n$, these equations take the following forms \citep{MSS}:
\bes
\ben
\label{eq:gen1}
&& {\partial {\cal G}_\delta \over \partial \tau} + (1+{\cal G}_\theta){\cal G}_\delta=0; \\
&& {\partial {\cal G}_{\theta}\over \partial \tau} + {1\over 2}{\cal G}_{\theta}+{1\over 3}{\cal G}^2_{\theta}+ {\cal G}_{\nabla^2\Phi} = 0 ; \\
\label{eq:g2}
&& {\cal G}_{\nabla^2\phi} = {3 \over 2}{\cal G}_\delta.
\label{eq:g3}
\een
\ees
The solution to these equations are well known and ${\cal G}_\delta$ and ${\cal G}_\theta$ satisfy collapse of spherically
over-dense top-hat perturbation. However, Eq.(\ref{eq:gen1})-Eq.(\ref{eq:g3}) do not represent evolution of perturbations
they encode the statistical description of an ensemble of perturbations.

To relate the tracer density $\delta_h$ with $\delta$ many different simplifying assumption
are employed. It is typically assumed the number density of tracers (proto-halos) do not change and remains
conserved. Thus evolution of $\delta_h$ can be described by a continuity equation. In this picture
the halos can change shape or their topology but they follow a well-defined trajectories.
It is further assumed that halo velocities  ${\bf v}_h$ are unbiased estimators of underlying dark matter velocities
${\bf v}$ i.e. ${\bf v}_h={\bf v}$:
\ben
(\dot\delta_h-\dot\delta) + \nabla_i [(\delta_h-\delta){v^i}]=0.
\label{eq:master}
\een
The overdots represent derivative w.r.t. $\tau$.
The halo density contrast $\delta_h({\bf x},\tau_i)$ and the DM density contrast $\delta({\bf x},\tau_i)$ 
are related at some initial time $\tau_i$ as follows:
\ben
&& \delta_h({\bf x},\tau_i)\equiv b(\delta) = \sum_{\ell} {b^L_\ell(\tau_i) \over \ell!} [\delta({\bf x},\tau_i)]^{\ell}
 = \sum_{\ell} {b^L_\ell(\tau) \over \ell!} [\delta({\bf x},\tau)]^{\ell}.
\een
Thus the evolution of the Lagrangian bias $b^L_{\ell}(\tau)$ as a function of conformal time $\tau$ from  $b^L_{\ell}(\tau_i)$ takes the following form:
$b^L_{\ell}(\tau) = b^L_{\ell}(\tau_i)\left [ {a(\tau_i)/a(\tau)} \right ]^{\ell}$.
The above expression can be used to evaluate $b^L_{\ell}$ at a later time $\tau$
once specified at an initial epoch $\tau_i$. Notice that at this stage we have left the parameters $b^L_{\ell}$ arbitrary.
Perturbative analysis of Eq.(\ref{eq:master}) has been carried out in an order-by-order manner.
In \citep{ChecnScocSheth,Baldauf} an analysis was performed up to second order in the linear density contrast $\delta^{(1)}$,
more recently the result was extended to third order in \citep{halo}. 
These studies found that the
halo density contrast $\delta_h$ is related to the underlying $\delta$ through the
following expression:
\bes
\ben
&& \delta_h \equiv b(\delta,\theta, \nabla_i\nabla_j\phi,\nabla_i\theta_j,\cdots );\\
&&\delta_h= b_1\delta + {1\over 2!}b_2\delta^2 + {1\over 3!}b_3\delta^3 +
{1\over 2!}b_{s^2} s^2 + b_{\psi}\psi + b_{st} s\cdot t  \nn\\
&&\quad\quad  + b_{\nabla^2\delta} \nabla^2\delta + b_{\nabla^2} \nabla^2[s_{ij}s^{ij}] + b_{\nabla^4} \nabla^2[s_{ij}]\nabla^2[s^{ij}] \cdots.
\label{eq:def_bias1}
\een
\ees
The coefficients $b_{s^2}, b_{\psi} \cdots$ and the higher-order derivative operators 
$b_{\nabla^2\delta}, b_{\nabla^2}, \cdots$ appearing in Eq.(\ref{eq:def_bias1}) are left arbitrary at this stage.
The non-local operators $t,s,\eta$ ans $\psi$ above are defined as \citep{Roy}:
\bes
\ben
\label{eq:def_s}
&& s_{ij}= {2\over 3{\cal H}^2}\nabla_i\nabla_j \phi -{1\over 3}\delta^{\rm K}_{ij} \delta;  \\
&& t_{ij}= \partial_i v_j -{1\over 3}\delta_{ij} \theta - s_{ij};  \\
\label{eq:def_t}
&& \eta = \theta -\delta; \\
\label{eq:eta}
&& \psi = \eta -{2\over 7}s^2 +{4\over 21}\delta^2.
\label{eq:psi}
\een
\ees
we have introduced the following notations
\ben
s^2 \equiv s_{ij}s^{ij}; \quad\quad s\cdot t \equiv s^{ij} t_{ij}; \quad\quad t^2\equiv t_{ij}t^{ij}.
\een
We have also assumed a $\Omega=1$ universe. However, the higher-order statistics are known to be very weakly-dependent
on background cosmology. The traceless tidal tensor is denoted as $s^{ij}$.
Here, $t_{ij}$ is considered to be symmetric as vorticity is not generated at lower-order in perturbation theory
only needs to be accounted for at a very higher-order. The terms $\eta$ and $t$ start to contribute at second-order
while $\psi$ contributes at cubic order and beyond. 

These operators along with density $\delta$ are invariants under the extended Lifshitz and Galilean transformation.
The local bias expansion corresponds to the invariant $\delta$
and represents a Taylor expansion of $\delta_h$ with coefficients $b_{\ell}$ specifying the exact 
functional form of $b(\delta)$. However, this is incomplete as inherent 
extended Lifshitz and Galilean symmetry of the Euler-continuity-Poisson system also
allows the additional invariants $s^2,t^2$ and $s\cdot t$ involving $s^{ij}$ and $t^{ij}$ etc. 
It has been argued that even if these forms of bias are not present in the initial conditions
there is no guarantee that they will not be generated during the subsequent gravitational evolution
as they are permitted by the symmetry of the system. It is expected that on large-scales  the polynomial model will be more accurate.
The non-local derivative terms in Eq.(\ref{eq:def_bias1}) are an unavoidable consequence of symmetry and will contribute on smaller scales.
Indeed, modification of gravity doesn't necessarily respect the symmetry under these transformations,
and, hence, in addition to non-local terms, scale-dependent terms will also be generated.

We can use Eq.(\ref{eq:def_bias1}) to relate the generating function  ${\cG}_{\delta_h}$ of $\delta_h$
in terms of the generating function of other variables. 
The generating functions are defined in Eq.(\ref{eq:gen}).
Using Eq.(\ref{eq:gen_1}) we arrive at:
\ben
&&{\cG}_{\delta_h}\equiv  b(\cG_{\delta})= b_1\cG_{\delta} + {1\over 2!}b_2[\cG_{\delta}]^2 + {1\over 3!}b_3[\cG_\delta]^3 +
{1\over 2!}b_{s^2} {\cG_s^2} + b_{\psi}{\cG_\psi} + b_{st} \cG_{s\cdot t} + \nn \\
&&\quad\quad\quad  + b_{\nabla^2\delta} {\cal G}_{\nabla^2\delta} + b_{s^2\nabla^2} {\cal G}_{s^2\nabla^2}  +  
b_{s^2\nabla^4}{\cal G}_{s\nabla^4} \cdots.
\label{eq:series}
\een
This is one of the main result of this paper. Following the derivations outlined in Appendix-\ref{sec:append_sym}  it can be shown that many of the terms
involving the following generating functions vanish.
\ben
{\cal G}_{s^2} =0; \quad {\cal G}_{t^2}=0; \quad {\cal G}_{s\cdot t} =0; \quad {\cal G}_{s^2\nabla^2}=0; \quad {\cal G}_{s\nabla^4}=0.
\een
An important conclusion from this analysis is that the higher-order statistics of tracers are independent
of $b_{s^2}, b_{t^2}, b_{st}$ and other similar constructs to an {\em arbitrary order} though they do contribute to the variance.

These quantities are the well known invariants that are the result of  
inherent extended Galilean and Lifshitz symmetries in the dynamic equations\citep{ChecnScocSheth,Baldauf,Kehagias}.
The corresponding expressions in terms of the generating functions take the following form:
\ben
\cG_{\eta} = \cG_{\theta}-\cG_{\delta}; \quad \cG_{\psi} = \cG_{\eta} -{2 \over 7}\cG_{s^2}
+{4 \over 21} [\cG_{\delta}]^2.
\een
It is possible to show using the properties of the generating functions in Eq.(\ref{eq:gen_1})-Eq.(\ref{eq:gen_n}) we have:
\ben
\cG_{\psi} = {\cal G}_{\eta} - {4\over 21}[\cG_{\delta}]^2.
\een
The Eulerian bias $b_{\ell}$ and the Lagrangian bias $b_{\ell}^L$
are related by the following expression \citep{halo}:  
\ben
b_1 = 1+ b_1^L;\quad b_2 = b_2^L + {8\over 21}b_1^L ;\quad b_{s^2} =-{4\over 7}b_1^L; 
\quad b_{\psi}=-{1\over 2}b_1^L; \quad b_{s t}=-{5\over 7}b_1^L; 
\label{eq:zero}
\een
It is recognized that galaxy formation is a stochastic process \citep{DekelLahav}.
A more general expression of Eq.(\ref{eq:def_bias1}) should include the stochasticity of galaxy formation with $\delta$ replaced by $\delta + n$ with
$n$ given by a more generic series expansion:
\bes
\ben
\label{eq:noise}
&& n= b_{\epsilon}\epsilon + b_{\delta\epsilon} \delta\epsilon + {1\over 2}b_{\delta^2\epsilon} \delta^2\epsilon + {1\over 2} b_{s^2\epsilon}s^2 \epsilon
+ {1\over 2} b_{\epsilon^2} \epsilon^2 + {1\over 3} b_{\delta\epsilon^2} \delta\epsilon^2 +{1\over 3}b_{\epsilon^3}\epsilon^3 + \dots; \\
&& {\cal G}_n = b_\epsilon {\cal G}_{\epsilon} + b_{\delta\epsilon} {\cal G}_{\delta}{\cal G}_{\epsilon} +  
{1\over 2}b_{\delta^2\epsilon} {\cal G}^2_\delta{\cal G}_\epsilon +{1\over 2}b_{s^2\epsilon} {\cal G}_{s^2}{\cal G}_\epsilon +
{1\over 2}b_{\delta\epsilon^2}{\cal G}_{\delta}{\cal G}^2_{\epsilon}+ {1\over 2}b_{\epsilon^3}{\cal G}^3_{\epsilon}.
\label{eq:noise1}
\een
\ees
The generating function ${\cal G}_\delta$ in Eq.(\ref{eq:series}) will be replaced by ${\cal G}_{\delta+n} = {\cal G}_{\delta}+{\cal G}_n$
Indeed, following the same arguments ${\cal G}_{s^2}=0$ and rest of the terms can be expressed in terms of ${\cal G}_{\delta}$ and ${\cal G}_{\epsilon}$.
If we assume $\epsilon$ to be Gaussian the expressions can be further simplified. 

It is also possible to consider a biasing model where the halo over-density at a given location
is assumed to be a function of dark matter fields and their higher-order derivatives
along the entire past trajectory. Such an expression would not only be 
non-local in space but also in time. It can however be argued that dominant perturbative expressions
can be factorized in spatial and temporal dependence. The integration of the temporal part
can be performed without distorting the spatial dependence. Thus only the parameters
defining the bias will get normalized.

The Eulerian bias in Eq.(\ref{eq:zero}) was expressed in terms of Lagrangian bias using order-by-order perturbative
calculation. The subset of polynomial bias coefficients $b_\ell$ can also be derived using the following mapping
that relates the Eulerian density contrast $\delta^{E}_{h}(\tau)$ for halos with the
Lagrangian density contrast $\delta^{L}_{h}(\tau)$:
\bes
\ben
\label{eq:spherical1}
&& 1+\delta^{E}_{h}(\tau)= (1+\delta_{})[1+\delta^{L}_h(\tau)];\\
\label{eq:spherical2}
&& 1+{\cal G}^{(h)}_E = (1+{\cal G}_\delta)(1+{\cal G}^L_{\delta});\\
&& \delta^{E}_h({\bf x},\tau) = \sum^{\infty}_{\ell=1} 
{b^E_\ell \over \ell!}[\delta({\bf x}, \tau)]^\ell; \quad 
\delta^{L}_h({\bf x},\tau) = \sum_{\ell=1}^{\infty} {b^L_\ell \over \ell!}[\delta^L({\bf x},\tau)]^\ell; \quad \\
&& {\cal G}^h_L(\tau) = \sum^{\infty}_{\ell=1} b^L_\ell {[{\cal G}_{\delta}(\tau)]^\ell\over \ell!}; \quad 
{\cal G}^h_E(\tau) = \sum^{\infty}_{\ell=1} b^E_\ell {[{\cal G}^L_\delta(\tau)]^\ell\over \ell!}; 
\quad {\cal G}_{\delta}(\tau) = \sum^{\infty}_{\ell=1} {\nu_\ell \over \ell!}\tau^{\ell}.
\label{eq:spherical3}
\een
\ees
We also express the Lagrangian and Eulerian generating functions as:
\ben
{\cal G}^L_{\delta}(\tau)=\tau =\sum^{\infty}_{\ell=1} a_\ell^I [{\cal G}_\delta(\tau)]^\ell.
\een
The above expansion is an inverse series of ${\cal G}_\delta$.
Using Eq.(\ref{eq:spherical2}) in Eq.(\ref{eq:spherical3}) we arrive at the following relations \citep{HJM1,HM}:
\bes
\ben
\label{eq:b1}
&& b^E_1(\tau) = 1+ b_1^L(\tau);\\
&& b^E_2(\tau) = 2(1+a^{I}_2)b_1^L(\tau) + b_2^L(\tau);\\
&& b^E_3(\tau) = 6(a^{I}_2+a^{I}_3)b_1^L(\tau) + 3(1+2a^{I}_2)b_2^L(\tau) + b_3^L(\tau).
\label{eq:b3}
\een
\ees
However, we would like to point out that in our derivation we have not assumed a spherical collapse model at any stage.
The $a_n$ parameters above are related to the $\nu_n$ parameters defined before $a_n=n!\nu_n$ which are determined by solving the dynamical equations 
Eq.(\ref{eq:gen1})-Eq.(\ref{eq:g3}).
\ben
&& {a_n}= \left \{1, {17\over 21}, {341\over 567}, {55805\over 130977}, \cdots \right \}; \quad
{a^I_n}= \left \{1, -{17\over 21}, {2815\over 3969}, -{590725\over 916839}, \cdots \right \}.
\een
The coefficients ${a^I_n}$ are the coefficients of the inverse series.
Taylor expanding ${\cal G}_{\delta}^{\rm ZA}$ and  ${\cal G}_{\delta}^{\rm PZA}$ 
and replacing the $a_n$ coefficients in Eq.(\ref{eq:b1})-Eq.(\ref{eq:b3})
with the $\mu_n$ coefficients will produce the resulting $b_n(\tau)$ parameters for the Zel'dovich (ZA) or post Zel'dovich approximation (PZA)
(see Appendix-\ref{sec:append_LPT} for a detailed discussion).  
One important point is probably worth mentioning here. Unlike previous derivations, e.g. \citep{HJM1,HM},
the above derivation is directly derived from
of Euler, Continuity and Poisson given in Eq.(\ref{eq:gen1})-Eq.(\ref{eq:g3}).

Thus at this level, all the coefficients that describe the generating function of the so-called {\em proto-halos} 
defined in Eq.(\ref{eq:series}) can be expressed in terms of the coefficients 
$b_1^L$ and $b_2^L$. These coefficients can be derived using a halo model based approach \citep{Saito}:

An important conclusion of this section would thus be that in a non-local bias model the
clustering of halos only depend on the local Lagrangian bias parameters $b^L_{\ell}$ and clustering
of density $\delta$  and the divergence velocity field $\theta$ 
that are characterized by the
vertices $\nu_n$ or $\mu_n$. 
This extends the result presented in ref.\citep{Munshi_IBIT}.
Next we will consider the case of scale-dependent bias.
%
%
%
%
%
%
%

%
\section{Scale-Dependent Bias}
\label{sec:bias}
%
%
The models discussed in \textsection\ref{sec:non} are not scale-dependent as the parameters $b^L_\ell$ are
independent of the wavenumbers $\bq_i$. 
The formalism of scale-dependent bias was developed in a series of papers: \citep{Matsubara,jack1,jack2,verde}
\ben
&& \delta_h(\bk) = \sum^{\infty}_{n=1} {1\over n!}\int{d^3\bq_1 \over (2\pi)^3}\cdots \int{d^3\bq_n \over (2\pi)^3} 
c^L_n(\bq_1,\dots,\bq_n)\delta_L(\bq_1)\cdots\delta_L(\bq_n)\delta_{\rm D}(\bk-\bq_{1\dots n}); \nn \\
&& \bq_{1\dots n} \equiv \bq_1+\cdots+\bq_n.
\een
Here $\delta_L$ is the linear density contrast in a perturbative expansion.  
In the Fourier domain the scale-dependent bias is implemented by 
replacing the scale-independent $b^L_{\ell}(\tau)$ 
parameters defined in previous section with the following 
functions $c^L_{\ell}(\bk,\tau)$ of wave numbers \citep{jacques2}:
\ben
\label{eq:c1}
&& c^L_1(\bq,\tau_i) = b^{\rm L}_{10}(\tau_i)+ b^{\rm L}_{01}(\tau_i) \bq^2\\
&& c^L_2(\bq_1,\bq_2,\tau_i) = {b^L_{20}}(\tau_i) + b^L_{11}(\tau_i)(\bq_1^2+ \bq_2^2)\nn \\
&&\quad + b^L_{02}(\tau_i)\bq_1^2\bq_2^2 -2 \chi^{L}_{10}(\bq_1\cdot\bq_2) + \chi^L_{01}(\tau_i)
\left [ 3(\bq_1\cdot\bq_2)^2 - \bq_1^2\bq_2^2\right ]
\label{eq:c2}
\een
An angular averaging of $c^L_n$ recovers the scale-dependent parameters $b^{L}_{\ell}$.
The functions $c^L_1$ and $c^L_2$ which depend on the parameters $b^L_{ij}$, $\chi^L_{ij}$ 
can be computed using peak-background split in \citep{vincent,jacques2}.
The form of the bias functions is based on rotationally symmetric invariants.
The peaks of the smoothed density fields are defined up to second order in derivatives
which explains the absence of terms with higher powers in $k$.
The Zel'dovich approximation (ZA) is used to map the Lagrangian positions to Eulerian position. 
In the real or configuration space \citep{jacques2}:
\ben
&& \delta_h\xti = b^L_{10}(\tau_i)\delta\xti-b^L_{01}(\tau_i)\Delta \delta\xti \nn \\
&& \quad\quad + {1\over 2!}b^L_{20}(\ti)[\delta\xti]^2 - b^L_{11}(\ti) \delta\xti\Delta\delta\xti
+{1\over 2} b^L_{02}(\ti)[\Delta\delta\xti]^2 \nn \\
&& \quad\quad + \chi^L_{10}(\tau_i)\, \nabla\delta\xt\cdot\nabla\delta\xt 
+ {1\over 2!}\chi^L_{01}(\tau_i)\left[ 3\nabla_i\nabla_j\delta -\delta^{(\rm K)}_{ij}\Delta\delta \right ]^2
+\cdots
\een
In terms of $\psi$, $\eta$ defined before \citep{jacques2}:
\ben
&& \delta_h\xt = b_{10}\delta\xt -b_{01}\Delta\delta\xt + {1\over 2!} b_{20} \delta^2\xt \nn\\
&& \quad\quad + {1\over 2!} b_{s^2} s^2\xt + b_{\psi}\psi\xt + b_{st} s\xt\cdot t\xt + \cdots
\een
The generating functions of $\delta_h$ and $\delta$ are related by the following expression:
\ben
&& \cG_{\delta_h} = b_{10}\cG_{\delta} -b_{01}\cG_{\Delta\delta} + {1\over 2!}b_{20}\cG_{\delta}^2 + {1\over 2!} b_{s^2}\cG_{s^2} \nn \\
&& \quad\quad + {1\over 2}b_{02}[{\cal G}_{\triangle \delta}]^2- b_{11}{\cal G}_{\delta}{\cal G}_{\triangle\delta} + b_{\psi}\cG_\psi + b_{st}\; \cG_{s\cdot t} + \cdots
\een
Thus the generating function $\cG_{\delta_h}$ at second order is determined by $\cG_{\delta}$ and $\cG_{\triangle\delta}$.
The following expressions relate the Eulerian bias coefficients $b_{ij}$ with their Lagrangian counterparts $b^L_{ij}$:
\ben
&& b_{10} = 1+ b_{10}^L;\quad b_{01} = -R_v^2+ b_{01}^L; \quad b_{20} = b_{20}^L + {8\over 21}b_{10}^L ; \nn \\
&& b_{s^2} =-{4\over 7}b_{10}^L;  \quad b_{\psi}=-{1\over 2}b_{10}^L; \quad b_{s t}=-{5\over 7}b_{10}^L;
\label{eq:bias_coeff}
\een
The expressions in Eq.(\ref{eq:zero}) gives statistics of $\delta_h$
in terms of the coefficients $b_{s^2}$, $b_{\psi}$ and $b_{st}$.
In addition to ${\cal G}_{\delta}$ it also depends on ${\cal G}_{\triangle \delta}$.
Scale dependent bias has also been used in the context of primordial non-Gaussianity \citep{Desjacques} which we have
ignored here. However, the results discussed here can trivially extended to include primordial non-Gaussianity.
%

\section{Cumulants and Cumlant Correlators}
\label{sec:cumu}
In this section we use the results derived in previous
sections to compute the higher-order one-point and two-point
statistics of biased tracers.
The cumulants and their correlators for the halos and the underlying dark matter distribution is defined as follows:
\ben
{\cal S}_{\rm N}^{(h)} = {\la \delta_h^{\rm N}\ra_c \over \la \delta^2_h\ra_c^{\rm N-1}}; \quad 
{\cal C}_{\rm pq}^{(h)} = {\la \delta_{h1}^{p}\delta_{h2}^q\ra_c \over \la\delta_{h1}\delta_{h2} \ra_c \la \delta^2_h\ra_c^{\rm p+q-2}}; 
\quad \delta_{hi}\equiv \delta({\bf x}_i).
\label{eq:cc}
\een
A similar expression holds for the underlying dark matter distribution and will be denoted without the subscript ${}_h$ \citep{FryGaztanga}.
\ben
\label{eq:cc1}
\nu_1^{(h)} = b_1; \quad
\nu_2^{(h)} = (b_2 + b_1 \nu_2); \quad 
\nu_3^{(h)} = (b_3 + 3 b_2 \nu_2 + b_1 \nu_3).
\label{eq:cc4}
\een
For $b_1=1$ and $b_n=0$ we recover the unbiased result $\nu_n^{(h)} = \nu_n$.
In practice the $b_n$ are computed  using the Press-Schechter (PS) or Sheth-Tormen (ST) 
mass functions or using theories based on peak statistics.
The expressions of $b_n$ are given in Eq.(\ref{eq:zero}) and Eq.(\ref{eq:bias_coeff}).
The cumulants ${\rm S}_n^{(h)}$ can be expressed in terms of the vertices $\nu^{(h)}_n$ \citep{Fry,Ber92}:
\ben
\label{eq:S_N}
{\cal S}^{(h)}_3= 3\nu^{(h)}_2; \quad
{\cal S}^{(h)}_4 = 4\nu^{(h)}_3 + 12[\nu^{(h)}_2]^2.
\label{eq:S_Np}
\een
In the perturbative regime the following relations hold \cite{Ber92,Ber94}:
\ben
{\cal S}_3 = {34\over 7} + \gamma_1;\quad
{\cal S}_4 = {60712 \over 1323} + {62 \over 3}\gamma_1 + {7 \over 3}\gamma_1^2 +{2 \over 3}\gamma_2.
\een
The terms involving $\gamma_p=[{d^p \log\,\sigma^2(R_0)/d(\log R_0)^p}]$ are results of smoothing using top-hat window
of radius $R_0$.
We will use the following notation to represent the variance of the smoothed field $\sigma^2(R_0) =\la\delta_s^2\ra_c$ and 
correlation function $\xi_{12}=\la\delta_s({\bf x}_1)\delta_s({\bf x}_2)\ra_c$. 
The CCs take the following form:
\ben
\label{eq:c21_DM}
{\cal C}^{(h)}_{21} = 2\nu^{(h)}_2; \quad {\cal C}^{(h)}_{31} = 3\nu^{(h)}_3 + 6\nu^{(h)}_2; 
\label{eq:c51}
\een
The CCs satisfy a factorization property in the large-separation limit [$\xi_{12}(|{\bf x}_1 -{\bf x}_2|) < \sigma^2(R_0)$]:
${\rm C}^{(h)}_{pq} = {\rm C}^{(h)}_{p1}{\rm C}^{(h)}_{q1}$. Here, $\sigma^2(R_0)$ is the variance of the
smoothed density field, and $\xi_2(|{\bf x}_1 -{\bf x}_2|$ represents the two-point correlation function.
A tophat smoothing window with a radius $R_0$ is assumed.
In the quasi-linear regime with a tophat smoothing window the CCs have the following expressions \citep{francis}: 
\ben
\label{eq:c21}
{\cal C}_{21} = {68\over 21}+ {\gamma_1 \over 3}; \quad
{\cal C}_{31} = {11710 \over 441} + {61 \over 7}\gamma_1 + {2 \over 3}\gamma_1^3 +{\gamma_2\over 3}.
\label{eq:c41}
\een

The cumulant correlators probe squeezed and collapsed configuration of the underlying multispectra.
The related statistics in the Fourier domain are the squeezed bispectrum and the squeezed and collapsed
trispectrum. The lowest order non-trivial ${\cal C}_{21}$ is independent of the contribution from $s_{ij}s^{ij}$.
In a similar manner the squeezed bispectrum do not take any contribution from $s_{ij}s^{ij}$ in Fourier
domain. It is expected that similar results will hold for the squeezed and collapsed
trispectrum, ${\cal C}_{31}$ and ${\cal C}_{22}$ respectively.
In \textsection\ref{sec:non} we have shown that all additional terms vanish and only contributions from $\psi$
need to be included. Taylor expanding ${\cal G}_{\psi}(\tau_s)$:
\ben
{\cal G}_{\psi}(\tau_s) =&& {4\over 21}+ {8\over 21}\tau_s +
{1\over 42}(8 + 21\mu_2  -13\nu_2)\tau_s^2 \nn \\
&& + {1\over 126} \left ( 21\mu_3 +24\nu_2 -13 \nu_3\right )\tau_s^3+ \cdots
\een
The expressions for $\nu^{(h)}_k$ defined in Eq.(\ref{eq:cc4}) now get modified and depend also on the bias coefficient $b_{\psi}=-b_L/2$ as:
\bes
\ben
\label{eq:cc1}
&& \nu_1^{(h)} = b_1 + {4\over 21}b_{\psi}; \quad\\
&& \nu_2^{(h)} = (b_2 + b_1 \nu_2) + {8\over 21}b_{\psi}; \quad \\
&& \nu_3^{(h)} = (b_3 + 3 b_2 \nu_2 + b_1 \nu_3) + {1\over 84}(8 + 21\mu_2  -13\nu_2)b_{\psi}.
\label{eq:cc4}
\een
\ees
Notice that ${\cal G}_{\eta}(\tau_s)$ takes contribution from both $\delta$ and $\theta$ vertices
thus making the $\delta_h$ statistics a function of both $\delta$ and $\theta$ Eq.(\ref{eq:cc1})-Eq.(\ref{eq:cc4}).
Indeed, Eq.(\ref{eq:noise})-Eq.(\ref{eq:noise1}) provide a framework for inclusion of arbitrary noise contribution.
In case of a Gaussian noise, the higher-order terms of Eq.(\ref{eq:noise1}) will not contribute and only the variance 
will be affected through the term $b_{\epsilon^2}\epsilon^2/2$. 

For two different populations of tracers $h$ and $h^{\prime}$ the CCs defined in Eq.(\ref{eq:cc}) can be generalized to:
\ben
{\cal C}_{pq}^{(hh^{\prime})} = 
{\la \delta_{h1}^{p}[\delta_{h^{\prime}2}]^q\ra_c \over \la\delta_{h1}\delta_{h\prime2} 
\ra_c \la \delta^2_{h1}\ra_c^{\rm p-1}\la [\delta_{h\prime2}]^2\ra_c^{\rm q-1}}; 
\een
We have used the following notations $\delta_{h1}\equiv\delta_h({\bf x}_1)$ and $\delta_{h^{\prime}2}\equiv\delta_{h^\prime}({\bf x}_2)$,
the respective CCs can be factorized and be expressed in terms of respective CCs i.e. 
${\cal C}^{hh\prime}_{pq}= {\cal C}^{h}_{p1} {\cal C}^{h\prime}_{q1}$.
The CCs  ${\cal C}^{h}_{p1}$ and ${\cal C}^{h\prime}_{q1}$ are constructed from the coefficients of the
series expansion of their respective density contrasts as in Eq.(\ref{eq:series}). 
%
%
%
%
%
%
 
\section{Summary and Outlook}
\label{sec:conclu}
In a generic biasing scheme, based on symmetry arguments, many additional terms can be included.
Using a generating function formalism we test which of these terms actually contribute.
In \citep{Kehagias} symmetry arguments were used to determine the temporal dependence of the terms 
included in Eq.(\ref{eq:def_bias1}). Here, we use the symmetry to determine which terms will actually contribute.
Many other terms can be simplified drastically. We use the results to compute the higher-order one-point cumulants 
as well two-point CCs for collapsed objects.

The degeneracies present in the characterization of bias at the level of power spectrum can only be broken by
using higher-order statistics of biased tracers. 
In this paper we compute higher-order statistics in terms of the recently introduced {\em invariant}
quantities. The inherent symmetries of the dynamic
equations predict invariant quantities \citep{ChecnScocSheth,Baldauf,Kehagias} in the perturbative
expansion of bias. Using a generating function approach we use the
functional relation of $\delta_h$ in terms of $\delta$ and $\theta$ to relate the
tree-level vertices of $\delta_h$ with that of $\delta$ and $\theta$.
We derive formal relations in the presence of higher-derivative terms.
We have used these expressions to show that certain terms in this expression
do not contribute at any given order. The results from the generating function formalism 
is used to express the cumulants and CCs of biased tracers.
The squeezed bispectrum as well as the squeezed and collapsed tri-spectra are related to their counterparts of same order, respectively
to ${\cal C}_{21}$ and ${\cal C}_{31}$, ${\cal C}_{22}$. The results
developed here generalize the expressions derived for CCs in ref.\citep{Munshi_IBIT} assuming a 
polynomial bias model. Similar generalizations are possible for squeezed and collapsed multispectra.

classical physics at a fundamental level is deterministic. However, the detailed microscopic physics of 
galaxy formation is not well understood. A stochastic contribution $\epsilon$ is 
thus often included in the expression of bias to encode our lack of knowledge that relates $\delta_h$ with $\delta$ and $\theta$.
The terms in Eq.(\ref{eq:def_bias1}) correspond to  $\epsilon=0$ (no-noise). The terms in Eq.(\ref{eq:noise}) depicts the
first- and second-order terms in noise in a Taylor expansion of $\delta_h$ in terms of $\epsilon$.
In computing the higher-order statistics of any biased tracers these contributions
should be included. We have considered higher-order terms in $\epsilon$ and that represent its higher-order correlation with
$\delta$ and $\theta$. The generating function
approach is generalized to take in account presence of such stochastic contributions at an arbitrary order.
The generating function approach simplifies the order-by-order analysis.
Assuming a Gaussian stochastic noise can further simplify the expression as all terms beyond
second-order that characterize non-Gaussianity vanish. 

In recent years the large deviation principle (LDP) has been used to construct the one- and two-point 
PDFs of biased tracers \citep{Cora1,Cora2,Cora3} (also see \citep{valageas} for related approach based on steepest descent method).
This method is also related  to earlier generating function based approaches \citep{Ber92,Ber94,francis}.
Recent work based on LDP also has attempted to compute the PDF and bias of collapsed objects.,
These results were obtained assuming a polynomial biasing model.
Results presented here will help us to go beyond the polynomial model and include 
stochastic noise within the LDP formalism (Munshi 2017; in preparation).
%
  
%
%
\section*{Acknowledgments}
\label{sec:acknow}
The author acknowledges support from the Science and Technology
Facilities Council (grant numbers ST/L000652/1). 
It is a pleasure for the author to acknowledges useful discussion with members of the
University of Sussex cosmology group. The author would
like to thank Donough Regan for his help and suggestions to improve the draft.
\bibliography{bias.bbl}
\appendix
\section{Symmetries and Generating Function}
\label{sec:append_sym} 
In this appendix we provide a detailed derivation of of ${\cal G}_{s^2}=0$. These results are a direct 
consequence of the fact that $s_{ij}$ is a traceless tensor and the generating functions
which encode tree-level amplitudes of vertices in the perturbative 
analysis of Euler-Continuity-Poisson system satisfies spherical top-hat collapse equations;  in
a spherically symmetric setup the off-diagonal terms that are related to departure from spherical symmetry do not contribute.
We start with the definition of $s_{ij}$ in Eq.(\ref{eq:def_s}) which gives:
\ben
&& s_{ij}s^{ij} = {2 \over 3 {\cal H}^2} \nabla_{i}\nabla_j \phi\nabla_{i}\nabla_j \phi
-{2\over 3{\cal H}^2} \nabla_{i}\nabla_j \phi\, \delta^{K}_{ij} + {1 \over 9 }\,\delta^{K}_{ij}\,\delta^{K,ij}\delta^2.
\label{eq:sij_square}
\een
Summation over repeated indices is assumed. Next, we can simplify the first term using Eq.(\ref{eq:gen_3}) as:
\ben
{\cal G}_{\nabla_{i}\nabla_j \phi\nabla_{i}\nabla_j \phi} = {1\over 3}{\cal G}^2_{\delta}
\een
Using this expression in Eq.(\ref{eq:sij_square}) we arrive at the desired result.
A similar calculation can be used to prove ${\cal G}_{t^2}=0$ as well as ${\cal G}_{st}=0$.
These results are valid in the perturbative regime. Thus, it depends on
the assumption that the fluid flow is single stream and irrotational. 

Next we consider the terms involving the derivatives of $s_{ij}$ e.g. $\nabla^2[s_{ij}s^{ij}] $.
We note that $\nabla^2[s_{ij}s^{ij}] = 2\nabla^2[s_{ij}]s^{ij}$. So, we can write:
\ben
\nabla^2[s_{ij}] = \nabla_{i}\nabla_j\delta -{1\over 3}\delta^{K}_{ij}\nabla^2\delta
\een
Using Eq.(\ref{eq:gen_3}) as before we can write:
\ben
{\cal G}_{\nabla^2{s_{ij}s^{ij}}} = 2{\cal G}_{s_{ij}\nabla^2{s_{ij}}} = 0.
\een
The result ${\cal G}_{\nabla^2{s_{ij}}\nabla^2{s^{ij}}}=0$ can be derived using similar steps.
Following similar arguments we can prove similar identities for the divergence of velocity
in case of potential flow. These terms are not included in our definition of bias.  
%
%
\section{Bias and Lagrangian perturbation theory}
\label{sec:append_LPT}
It is possible to consider the Lagrangian perturbation theory (LPT) to model the underlying dynamics.
The Zel'dovich approximation (ZA) is first order in LPT. We list below
the generating functions at various order \citep{MSS}:
\ben
&& 1+ {\cal G}^{\rm ZA}_{\delta}(\tau_s) = \sum^{\infty}_{n=1} {{\mu}^{\rm ZA}_n\over n!}\tau_s^n =
 \left ( 1- {\tau_s \over 3}\right )^{-3}; \nn \\
&& 1+ {\cal G}^{\rm PZA}_{\delta}(\tau_s) =\sum^{\infty}_{n=1} {{\mu}^{\rm PZA}_n\over n!}\tau_s^n = \left ( 1- {\tau_s \over 3}-{\tau_s^2 \over 21}\right )^{-3}.
\een
Here $\rm PZA$ is the post Zel'dovich Approximation. A systematic development of higher order LPT in the 
context of generating function was developed in \citep{MSS}. For the Zel'dovich Approximation (1st order in LPT):
\ben
\{a_i\}^{\rm ZA} = \Big \{{2\over 3}, {10\over 27}, {5\over 27}, \cdots \Big \};  \quad 
\{a^I_i\}^{\rm ZA} = \Big \{-{2\over 3}, {14\over 27}, -{35\over 81}, \cdots \Big \}.
\een
The corresponding relation between Lagrangian and Eulerian bias are:
\ben
&& b_2^E = {2 \over 3}b_1^L + b_2^L; \quad b_3^E = -{16\over 9}b_1^L -b^L_2 +b^L_3.
\label{eq:LE_ZA}
\een
For PZA:
\ben
\{a_i\}^{\rm PZA} = \Big \{{17\over 21}, {106\over 189}, {47\over 1323}, \cdots \Big \};  
\quad \{a^I_i\}^{\rm PZA} = \Big \{-{17\over 3}, {992\over 1323}, -{20558\over 27783}, \cdots \Big \}.
\een
The corresponding relations between Lagrangian and Eulerian bias get modified to: 
\ben
&& b_2^E = {8 \over 21}b_1^L + b_2^L; \quad  b_3^E = -{158\over 441}b_1^L -{13 \over 7}b^L_2 +b^L_3.
\label{eq:LE_PZA}
\een
Eq.(\ref{eq:LE_ZA}) and Eq.(\ref{eq:LE_PZA}) are Lagrangian approximations to the exact expression in Eq.(\ref{eq:b3}).
These results can be trivially extended to expressions linking higher-order Lagrangian and Eulerian 
bias parameters

\end{document}